\documentclass[usenatbib]{mn2e}
\usepackage{color}
\usepackage{natbib}
\usepackage{paralist}
\usepackage{graphicx}
\usepackage{epstopdf}
\usepackage{epsfig}
\usepackage{amsmath}
\usepackage{tablefootnote}

\usepackage{amssymb}
\usepackage{lipsum}
\usepackage{float}
\usepackage{array}
\usepackage{multirow}

\title[ALIPHATIC HYDROCARBON CONTENT OF THE INTERSTELLAR DUST]{ALIPHATIC HYDROCARBON CONTENT OF INTERSTELLAR DUST}
\author[B. G\"{u}nay, T. W. Schmidt, M. G. Burton, M. Af\c{s}ar, O. Krechkivska, K. Nauta, S. H. Kable, A. Rawal]
{B. G\"{u}nay $^{1}$\thanks{E-mail: burcu.gunay@ege.edu.com.tr (BG); timothy.schmidt@unsw.edu.au (TWS)
melike.afsar@ege. edu.tr (MA)},
T. W. Schmidt$^{2}$, M. G. Burton$^{3,4}$, M. Af\c{s}ar$^{1}$, O. Krechkivska$^{2}$,
 \newauthor K. Nauta$^{2}$, S. H. Kable$^{2}$, A. Rawal$^{5}$ \\
\\
$^{1}$Department of Astronomy and Space Sciences,
                 Ege University, 35100 Bornova, \.{I}zmir, Turkey;\\
$^{2}$School of Chemistry,
                 UNSW Sydney, NSW 2052, Australia;\\
$^{3}$School of Physics,
                 UNSW Sydney, NSW 2052, Australia;\\
$^{4}$Armagh Observatory and Planetarium,
                 College Hill, Armagh, BT61 9DG, Northern Ireland, UK;\\
$^{5}$Mark Wainwright Analytical Centre, UNSW Sydney, NSW 2052, Australia}

\begin{document}
\maketitle
\label{firstpage}
\begin{abstract}
In the interstellar medium, carbon is distributed between the gas and solid phases. However, while about half of the expected carbon abundance can be accounted for in the gas phase, there is considerable uncertainty as to the amount incorporated in interstellar dust.

The aliphatic component of the carbonaceous dust is of particular interest because it produces a significant 3.4\,$\mu$m absorption feature when viewed against a background radiation source. The optical depth of the 3.4\,$\mu$m absorption feature is related to the number of aliphatic carbon C-H bonds along the line of sight. It is possible to estimate the column density of carbon locked up in the aliphatic hydrocarbon component of interstellar dust from quantitative analysis of the 3.4\,$\mu$m interstellar absorption feature providing that the absorption coefficient of aliphatic hydrocarbons incorporated in the interstellar dust is known.

We generated laboratory analogues of interstellar dust by experimentally mimicking interstellar/circumstellar conditions. The resultant spectra of these dust analogues closely match those from astronomical observations. The measurements of the absorption coefficient of aliphatic hydrocarbons incorporated in the analogues were carried out by a procedure which combined FTIR and $^{13}$C NMR spectroscopies. The absorption coefficients obtained for both interstellar analogues were found to be in close agreement (4.76(8) $\times$ 10$^{-18}$ cm group$^{-1}$ and 4.69(14) $\times$  10$^{-18}$ cm group$^{-1}$), less than half those obtained in studies using small aliphatic molecules. The results thus obtained permit direct calibration of the astronomical observations, providing rigorous estimates of the amount of aliphatic carbon in the interstellar medium.
\end{abstract}

\begin{keywords}
Interstellar dust, carbonaceous dust, aliphatic carbon, carbon abundance, carbon crisis
\end{keywords}

\section{Introduction}\label{Section1}

	The evolution of our galaxy is driven by the cycle of material between the interstellar gas and stars.  The raw material for star birth is expelled from the previous generations of stars into the interstellar medium (ISM).  It is then incorporated into new stars as part of a continuous cycle of material, driven by immense energy flows originating from the stars.

	Carbon is the 4th most abundant element in the ISM. There is a rich carbon chemistry in the ISM due to its chemical versatility, able to bond through three different orbital hybridisations; sp$^{3}$, sp$^{2}$, and sp. These different types of hybrid orbitals lead to different bond structures, and therefore carbon can be found in four forms: aliphatic (alkane, sp$^3$), olefinic (alkene, sp$^2$), aromatic (sp$^2$) and alkyne (sp).
	
	Rich carbon chemistry starts in the envelopes of the massive evolved stars which can produce carbon in their cores. Simple carbon-containing molecules react to form larger molecules and grains in the circumstellar medium of these carbon stars. The carbon-rich gas and grains are expelled into the ISM \emph{via} stellar winds \citep{Henning2004, Contreras2013}. It is estimated that a considerable amount of carbon (up to 70\,\% of the total) may be found in the interstellar carbonaceous grains \citep{Sandford1991, VanDishoeck2014}.
	
	In the ISM, the carbon abundance includes the total carbon in both the gas and solid phase. It is given in terms of the C/H ratio in ppm\footnote{ppm: parts per million}. The total carbon abundance observed in the ISM should be in agreement with cosmic carbon abundance estimations. However, there appears to be a discrepancy between the total cosmic carbon abundance, as measured through absorption towards some stars, and the (lower) levels estimated via extinction measurements in interstellar dust, which we overview below.  This discrepancy has been termed the ``carbon crisis'' \cite{Kim1996,Dwek1997}.

The ISM carbon abundance derived from Solar abundances \citep{Grevesse1998,Asplund2005,Asplund2009}, and meteoritic/protosolar abundances \citep{Lodders2003} gives up to 270\,ppm carbon.  \cite{Snow1995} estimated a total abundance of $225\pm50$\,ppm, but more recently carbon abundances have been studied using B type stars \citep{Sofia2001,Przybilla2008} and young F, G disk stars \citep{Sofia2001}. The abundances in young stars such as these should match the observational abundances in the ISM. \cite{Sofia2001} found the total ISM carbon abundance around $358\pm82$\,ppm from the carbon abundance in young F, G type stars.

The gas phase abundances along various sight lines can be studied using atomic and molecular spectral lines \citep{Cardelli1996, Parvathi2012}. \cite{Cardelli1996} found that for sightlines toward six stars within 600\,pc of the sun, there was no dependence on either direction or physical conditions of the gas. They determined a gas phase abundance of $140\pm20$\,ppm. In the recent study of \cite{Parvathi2012}, the gas phase abundance was found to be inhomogeneous. They determined a maximum gas phase carbon abundance of 464$\pm$57\,ppm towards HD\,206773 and a minimum value of $69\pm21$\,ppm towards HD\,207198, using the H column densities derived from \cite{Cartledge2004, Cartledge2006}. In these studies, the carbon abundance in the solid phase is estimated as the shortfall between the gas phase abundance and the estimated total abundance. Towards HD\,207198, \cite{Parvathi2012} estimate as much as $395\pm61$\,ppm carbon resides in the dust, taking as their reference the $464\pm57$\,ppm measured towards HD\,206773. \cite{Cardelli1996} found the average carbon abundance in the solid phase to be between $50-150$\,ppm (using H column densities from \citealt{Bohlin1978} and \citealt{Diplas1994}), adopting $240\pm50$\,ppm as an intrinsic carbon abundance in the local ISM. Similarly, \cite{Snow1995} found only about 85\,ppm available for dust.
			
	The common direct method of tracing interstellar dust properties is to study the extinction of stellar light due to the combined effect of scattering and absorption by dust particles. The interstellar extinction curves covering the NIR to FUV regions of the spectrum give clues as to the size and chemical composition of the dust particles \citep{Cardelli1989, Fitzpatrick1999}. The UV extinction bump centred around 2175\,\AA  \citep{Stecher1965} is the strongest extinction feature. Besides the role of the grain size, it is thought that $\pi-\pi^{*}$ transitions of sp$^{2}$ carbon (graphitic carbon) incorporated in the interstellar grains are responsible for this UV absorption feature \citep{Mathis1977, Kwok2009}.

	Models based on the extinction curves \citep{Mathis1977, Kim1996} can be used to estimate elemental abundances in the dust. From dust models, as much as 300\,ppm carbon has been estimated to be found in the solid phase, making dust a significant reservoir for the element \citep{Mathis1977, Kim1996}. Adding this to the average 140\,ppm found in the gas phase brings about a total abundance of $\sim440$\,ppm, which is at the upper end of the $358\pm82$\,ppm from \cite{Sofia2001}, consistent with the $464\pm57$ of \cite{Parvathi2012} but totally inconsistent with the total abundance estimate of \cite{Snow1995}, $225\pm50$\,ppm. The discrepancy between the lower estimates of total carbon abundances and dust models is known as the ``carbon crisis'' \citep{Kim1996}.  However, the models based on UV extinction curves only indirectly estimate the amount of carbon in the solid phase \citep{Mishra2017}. A direct spectroscopic measurement is desirable.
	
	There are prominent spectral features of carbonaceous dust in spectra of the ISM in the infrared region. These absorption features are  3.28\,$\mu$m, 3.4\,$\mu$m, 5.87\,$\mu$m, 6.2\,$\mu$m, 6.85\,$\mu$m and 7.25\,$\mu$m \citep{Dartois2004}. The 3.4\,$\mu$m absorption feature is of particular interest since it is the more prominent and prevalent feature towards IR background radiation sources.

	The 3.4\,$\mu$m absorption feature has been extensively observed through several sightlines toward the Galactic Centre \citep{Willner1979, Wickramasinghe1980, McFadzean1989, Tielens1996, Sandford1991, Pendleton1994, Chiar2000, Chiar2002, Chiar2013}, Local ISM \citep{Butchart1986, Adamson1990, Sandford1991, Pendleton1994, Whittet1997}, planetary nebula \citep{Lequeux1990, Chiar1998} and the ISM of other galaxies  \citep{Imanishi2000, Mason2004, Dartois2004, Geballe2009}.  The 3.4\,$\mu$m feature has also been detected in the spectra of Solar System materials such as meteorites (e.g. \citealt{Ehrenfreund1991}), Interplanetary Dust Particles (IDPs) (e.g. \citealt{Matrajt2005}) and cometary grains (e.g. \citealt{Sandford2006, MunozCaro2008}).

	The observed 3.4\,$\mu$m absorption is attributed to the aliphatic C-H stretch in carbonaceous dust.  Therefore, the optical depth of the 3.4\,$\mu$m  absorption feature is related to the number of aliphatic carbon C-H bonds along the line of sight. It is possible to estimate the column density, $N$, of carbon locked up in aliphatic hydrocarbon material of interstellar dust grains from quantitative analysis of the 3.4\,$\mu$m  interstellar absorption feature. One can measure the optical depth  ($\tau$) and the equivalent width ($\Delta \bar{\nu}$, cm$^{-1}$) of the 3.4\,$\mu$m  (2940 cm$^{-1}$) absorption and determine the number of $CH_x$ groups needed to produce this absorption by using the laboratory measurements of the integrated absorption coefficient ($A$, cm\,group$^{-1}$).
	
\[
N =\frac{\tau \Delta \bar{\nu}} {A}
\]

	To date several forms of hydrocarbon materials with different sp$^{2}$/sp$^{3}$ hybridisation and C/H ratios have been studied to determine the aliphatic integrated absorption coefficient, $A$, for astrophysical interest (e.g., \citealt{Hendecourt1986, Duley1998, Furton1999,  Mennella2002, Dartois2004, Steglich2013, Gadallah2015}). However, measurements have been carried out with small hydrocarbon molecules or ice residues, or random hydrocarbon materials for which the aliphatic content is only indirectly determined.  Interstellar dust analogues (ISDAs) have previously been produced in the laboratory, but the integrated absorption coefficient of the aliphatic component has not been rigorously obtained (\citealt{Mennella1999, Lee1993, Schnaiter1999, Kovacevic2005}).

There is a discrepancy between the results for small molecules, for which the structure is well defined \citep{Hendecourt1986,Dartois2004}, and for dust analogues, for which the structure is ill-defined \citep{Duley1981,Furton1999,Mennella2002}. The integrated absorption coefficient, per aliphatic group (one carbon atom), for the interstellar dust analogues is found to be less than half that of the small molecules. \cite{Duley1998} determined $A=4.0\times10^{-18}$\,cm\,group$^{-1}$ for CH$_3$ and $A=2.6\times10^{-18}$\,cm\,group$^{-1}$ for CH$_2$. This contrasts with the work of \cite{Dartois2004} who determined $A=14.5\times10^{-18}$\,cm\,group$^{-1}$ for CH$_3$ and $A=10.8\times10^{-18}$\,cm\,group$^{-1}$ for CH$_2$. Previous determinations of the integrated absorption coefficient are presented in Table \ref{tab:x}. For ease of comparison with the present results, it has been assumed that $N(CH_2)/N(CH_3)\approx2$, $x=2.33$ \citep{Dartois2007}.

\begin{table}
  \caption{Comparison of reports of integrated absorption coefficient of aliphatic dust analogues}
  \begin{tabular}{lrrr}
    \hline \hline
    sample & $A_{CH_x}$ & $\sigma_{CH_x}$  & $\kappa$ \\
   & (cm\,group$^{-1})$ & (cm$^2$\,group$^{-1}$) & (cm$^2$\,g$^{-1}$)\\
    \hline
$^a$ISDA-iso. & $4.76(8)\times10^{-18}$ & $4.26(7)\times10^{-20}$	& 1788(30)\\
$^a$ISDA-ac. & $4.69(14)\times10^{-18}$ & $4.19(13)\times10^{-20}$	& 1762(53)\\
\hline
S91$^b$	&$11.0\times10^{-18}$	\\			
D04$^b$	&$12.0\times10^{-18}$	\\
D98$^b$ & $3.1\times10^{-18}$\\
\hline
F99		& &	&				1000-1700\\
M02		& &	&				1600\\
G15		& &	&				1450-3345\\
    \hline
  \end{tabular}
$a$ This work.\\
  $b$ assuming $N(CH_2)/N(CH_3)=2$ \citep{Dartois2007}.\\
  \textit{x} = [($2\times N(CH_2) + 3\times N(CH_3))/(N(CH_2)+N(CH_3))$\\
 S91: \cite{Sandford1991}, D07: \cite{Dartois2007}, D98: \cite{Duley1998}, F99: \cite{Furton1999}, M02: \cite{Mennella2002}, G15: \cite{Gadallah2015}.\\
  \label{tab:x}
\end{table}

The problem with determining the integrated absorption coefficient of the aliphatic component of an interstellar dust analogues (ISDA) is one of determining the fraction of aliphatic carbon. A second spectroscopy is required to independently measure the aliphatic content of the sample.

For reliable quantitative analysis, the most direct measurement of the hydrocarbon composition of a sample is $^{13}$C NMR spectroscopy. The advantage of $^{13}$C  NMR spectroscopy is that each hybridisation gives rise to a separate signal with the same weighting factor \citep{Robertson2002}. Therefore $^{13}$C  NMR spectroscopy ensures the absolute determination of the amount of aliphatic carbon in the sample for quantitative purposes \citep{Henning2004}. However, until now, $^{13}$C NMR spectroscopy has not been applied to reliable interstellar dust analogues in order to determine the integrated absorption coefficient of the 3.4\,$\mu$m feature. Because the NMR technique requires a large sample, a long time is required to produce sufficient sample under low particle density conditions.

In this paper we report a holistic approach to determine the integrated absorption coefficient for ISDAs. We produce ISDAs in the laboratory under simulated circumstellar/interstellar-like conditions and find that their spectra closely match interstellar absorption profiles. Having established the similarity in absorption profile, the aliphatic content of the ISDAs is determined by $^{13}$C NMR spectroscopy. This information is used in concert with FTIR spectra to determine the integrated absorption coefficient, $A$, for aliphatic carbon in ISDAs. Combining $A$ with the optical thickness ($\tau_{3.4\,\mu m}$) allows one to estimate the column density of aliphatic hydrocarbon incorporated in the carbonaceous dust, which in turn a provides valuable approach to resolve the ``carbon crisis''.
	
\section{Experimental Methods}\label{Section2}
\subsection{Interstellar Dust Analogue Production}
The interstellar dust analogues were produced from acetylene (HC$\equiv$CH) and isoprene (H$_2$C=C(CH$_3$)-CH=CH$_2$). These samples are referred to here as ISDA-acetylene and ISDA-isoprene, respectively. The acetylene precursor was expected to generate a largely unsaturated ISDA, while the isoprene was used to favour a branched aliphatic structure.
	
The experimental apparatus used for interstellar dust analogue production consisted of a vacuum chamber (VC), and a diffusion pump, backed by a mechanical pump. The pressure inside the VC was around 10$^{-4}$ Torr during operation. The chamber was equipped with a pulsed discharge nozzle (PDN), operating on argon gas enriched with precursor molecules ($\sim1\,\%$ acetylene, 15\,\% isoprene). The duration and frequency of the nozzle pulse was $250 - 350$\,$\mu$s and $10 - 50$\,Hz. Each gas pulse was struck with an electrical discharge: A large negative voltage (2000\,V) was applied to one electrode, while the other electrode was held at the ground potential. The discharge plasma contains electrons, ions and metastable argon atoms which perform complex chemistry. The net result is the generation of plasma supersonically expanded into the vacuum chamber, which provides the relevant circumstellar dust formation conditions: $n = 10^{10} - 10^{12}$\,cm$^{-3}$, \cite{Contreras2013}. The resultant, condensed species (ISDA) accumulated in the collection zone of a clean Petri dish placed underneath the PDN.

\subsection{Analysis}
\subsubsection{Scanning Electron Microscopy}
The ISDA samples were investigated by scanning electron microscopy (SEM). The ISDAs were obtained by scraping the accumulated interstellar dust analogues from the collection zone surface, and depositing onto silicon substrates. Images were obtained using the Nova NanoSEM230 in the Mark Wainwright Analytical Centre of the University of New South Wales.

\subsubsection{UV Analysis}
UV spectra of the interstellar dust analogues were measured with Agilent Cary 100 UV-Vis Spectrometer. ISDA$-$isoprene and ISDA$-$acetylene samples were mixed with ethanol and exposed to ultrasonic waves in order to disperse them into their constituent particles. The ethanol was evaporated and the resultant samples were used to prepare suspensions in hexane.

\subsubsection{FTIR Analysis}
KBr (Sigma Aldrich $-$ FTIR Grade) was chosen as the matrix and substrate material owing to its transmission window in the IR region. KBr  was dried in an oven (24\,h, 200 $^{\circ}$C) to reduce spectral contamination due to adsorbed water. The sample pellets were prepared by diluting a known quantity of the ISDA in KBr, which was then pressed into a thin disk using a steel die (7\,mm) and a hydraulic press.

The absorption spectra of ISDAs were recorded with a VERTEX 70v FTIR spectrometer. All measurements were carried out in vacuum ($<0.2$\,mbar). FTIR spectral measurements were recorded for different column densities (cm$^{-2}$) of aliphatic carbon. ISDA-isoprene was studied with 20 sample pellets and ISDA-acetylene measurements were performed on 12 sample pellets. The background-subtracted, normalised FTIR spectra were used to obtain 3.4\,$\mu$m aliphatic C-H stretch absorption feature profiles. Total integrated areas of the resultant 3.4\,$\mu$m aliphatic absorption features were calculated to obtain the integrated absorbance, $\mathcal{A}$ (cm$^{-1}$) as a function of the columns density of aliphatic groups.

\subsubsection{Solid-state $^{13}$C NMR Measurements}
In order to find the proportion of aliphatic carbon atoms incorporated into the samples, we carried out quantitative $^{13}$C NMR measurements with a Bruker Avance III 300\,MHz Solid State NMR spectrometer. These measurements allowed us to calculate the weight ratio of the aliphatic carbon, $r_{C}$, in the pellets. The column density of the aliphatic carbon in the pellets was calculated using  $r_{C}$, the mass of the samples in the pellets, $m_{a}$, the molar mass of carbon, $M_{C}$, and the surface area of the pellets, $s_{p}$.

\[
N =N_A\frac{r_Cm_a}{M_Cs_p}
\]

\section{Results and Discussion}\label{Section3}
\subsection{Scanning Electron Microscopy}
\begin{figure}
  \leavevmode
    \centering
      \epsfxsize= 8cm
      \epsfbox{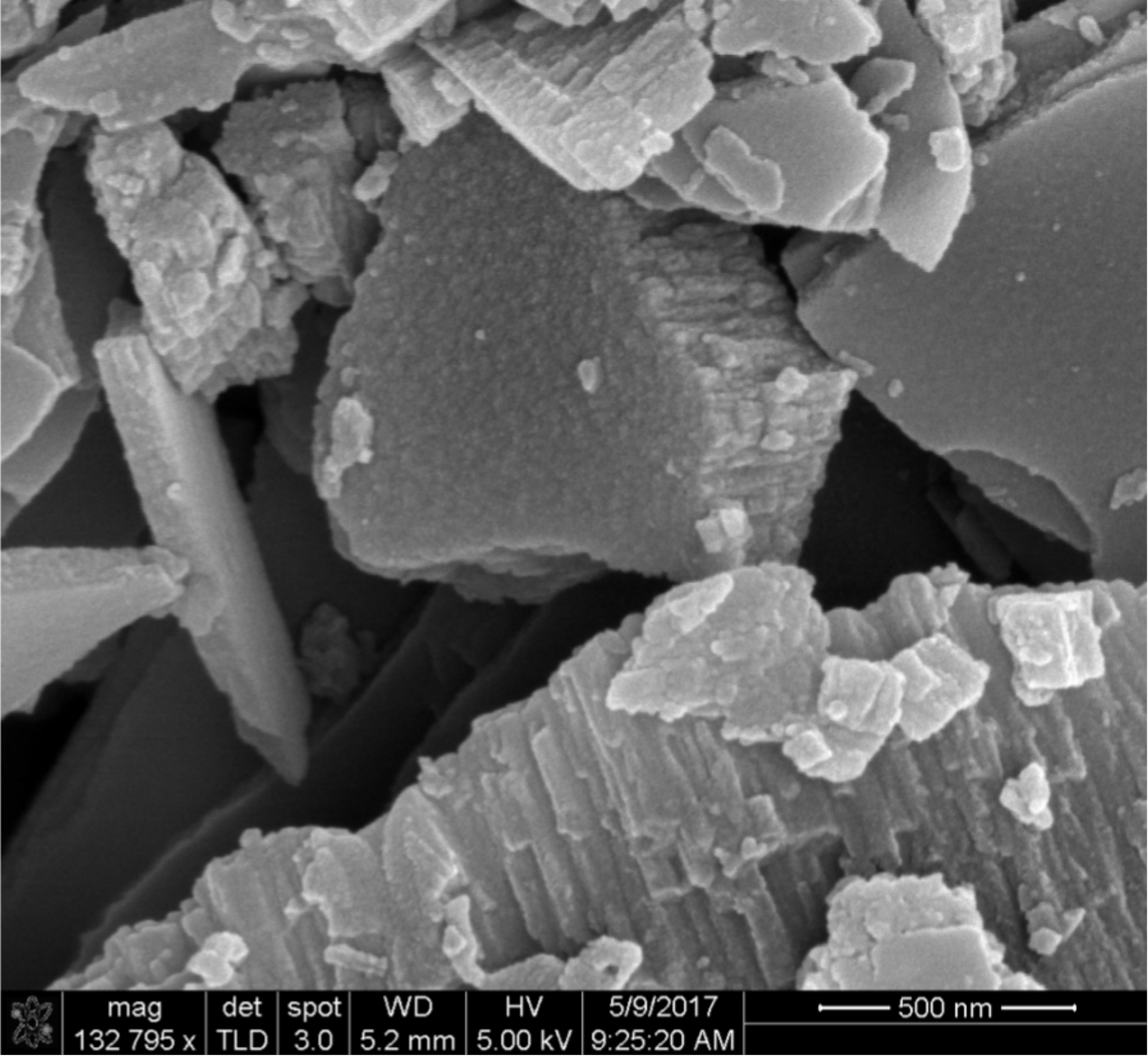}
       \caption{The SEM image of ISDA$-$acetylene showing sub-micron-sized graphitic particles.}
     \label{fig2}
\end{figure}

An SEM image of ISDA-acetylene is displayed in Figure \ref{fig2}. Since we could not collect an isolated ISDA particle, we cannot indicate their exact size. However, from the SEM image, it may be seen that the ISDA-acetylene consists of sub-micron-sized, substantially graphitic particles. These particles were found to agglomerate to form layers which fractured upon being scraped from the collection zone. The observation of planar, graphitic structures in our ISDA is consistent with the argument that the 2175\,\AA\ bump in the UV extinction curves is produced by the small interstellar graphite grains \citep{Kwok2009}. We could not obtain the surface structure of ISDA-isoprene in detail.

\subsection{UV Spectra}
\begin{figure}
  \centering
   \leavevmode
      \epsfxsize= 8cm
      \epsfbox{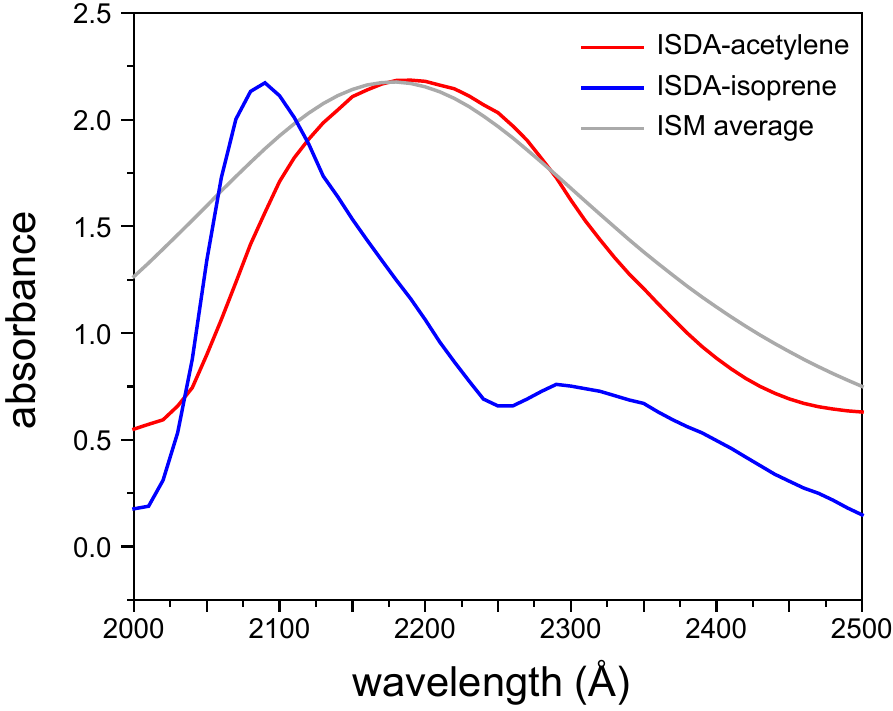}
       \caption{The UV absorption spectra of ISDA-acetylene and ISDA-isoprene suspended in hexane. The central UV absorption wavelength of ISDA-acetylene is at 2190\,\AA, which is close to the central wavelength of the observed UV extinction bump (from \citealt{Fitzpatrick2007}). }
     \label{fig6}
\end{figure}

Figure \ref{fig6} shows UV spectra of the ISDAs suspended in hexane. The central wavelength of UV absorption of ISDA$-$isoprene is observed somewhat shorter than 2100\,\AA. However, the central wavelength of UV absorption of ISDA$-$acetylene is at 2190\,\AA, which is close to the central wavelength of the UV extinction bump \citep{Fitzpatrick2007}. However, it is known that there are variations in the central wavelength of the interstellar feature (2175$\pm$9\,\AA, \cite{Mathis1994}). Thus, an absorption centred at 2190\,\AA\ can be considered close to the range of the UV extinction bump. However, there are also possible interactions with the solvent/medium molecules (hexane). Therefore, further analysis in vacuum or deposition of the samples on a UV-transparent substrate is required to precisely determine the central wavelength of the UV absorption of the ISDAs. Nevertheless, exhibition of a UV peak consistent with contributions to the UV extinction is consistent with our ISDAs resembling interstellar material.

\subsection{FTIR Measurements}
\begin{figure}
  \leavevmode
    \centering
      \epsfxsize= 8cm
      \epsfbox{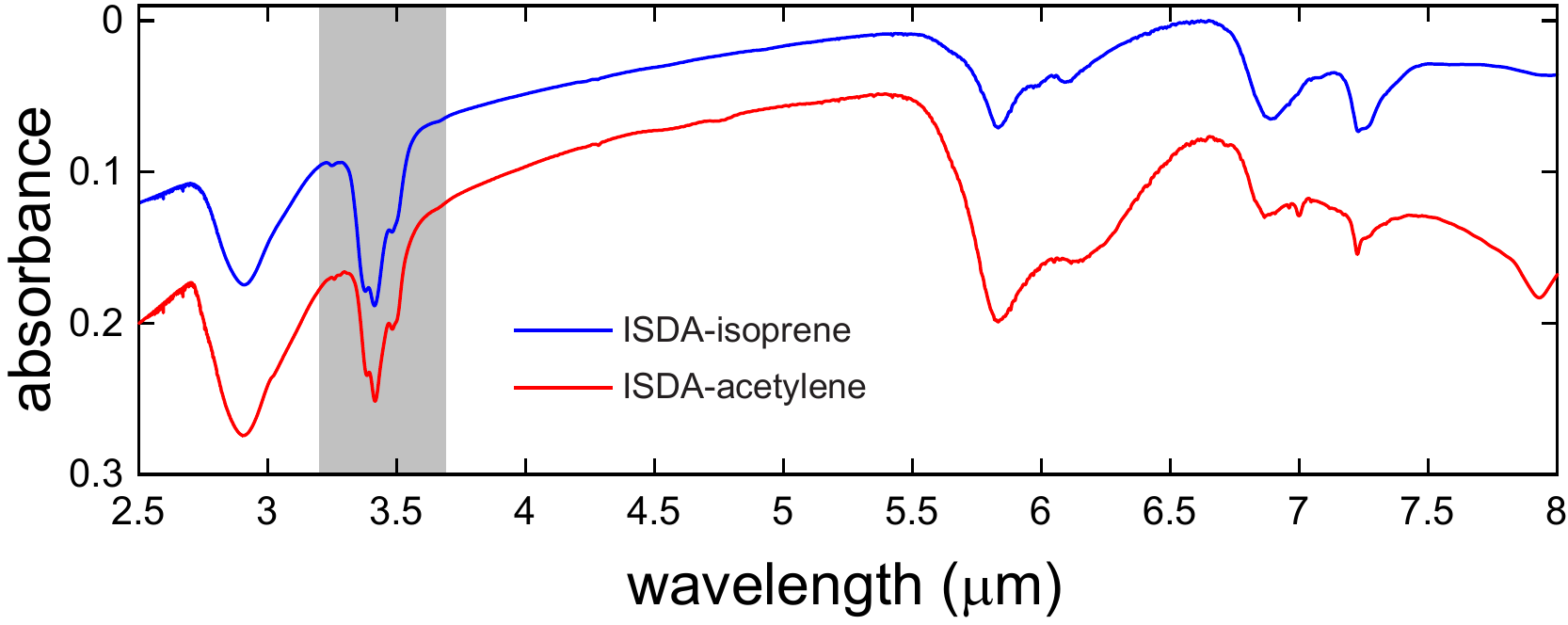}
       \caption{The raw mid-IR spectra of interstellar dust analogues: ISDA-isoprene and ISDA-acetylene (A normalising factor has been applied to facilitate the comparison). The aliphatic C-H stretching region has been shaded.}
     \label{fig3}
\end{figure}

The $2.5 - 8\,\mu$m range raw IR spectra of the ISDAs are presented in Figure \ref{fig3} (a normalising factor has been applied to facilitate the comparison). The IR spectra of ISDA$-$isoprene and ISDA$-$acetylene are found to be similar.

In the IR spectrum of the ISDAs, there are prominent absorption features of symmetric/asymmetric C-H stretching of CH$_{2}$ and CH$_{3}$ groups at 3.4\,$\mu$m with small features around 6.9\,$\mu$m and 7.25\,$\mu$m due to bending of CH$_{2}$ and CH$_{3}$ groups respectively. This demonstrates the presence of aliphatic material. There are additional features of aromatic and/or olefinic C-H and C=C stretches at 3.3\,$\mu$m and 6.2\,$\mu$m, suggesting a certain amount of aromatic/olefinic material. There are also carbonyl C=O stretching features around 5.8\,$\mu$m as the ISDAs were unavoidably exposed to the air. The hydroxyl O-H stretch arising from carboxylic acids and alcohols around 3.0\,$\mu$m could not be distinguished as this region is covered by broad O-H stretch feature due to atmospheric H$_{2}$O contamination of KBr.

\begin{figure}
  \leavevmode
    \centering
      \epsfxsize= 8cm
      \epsfbox{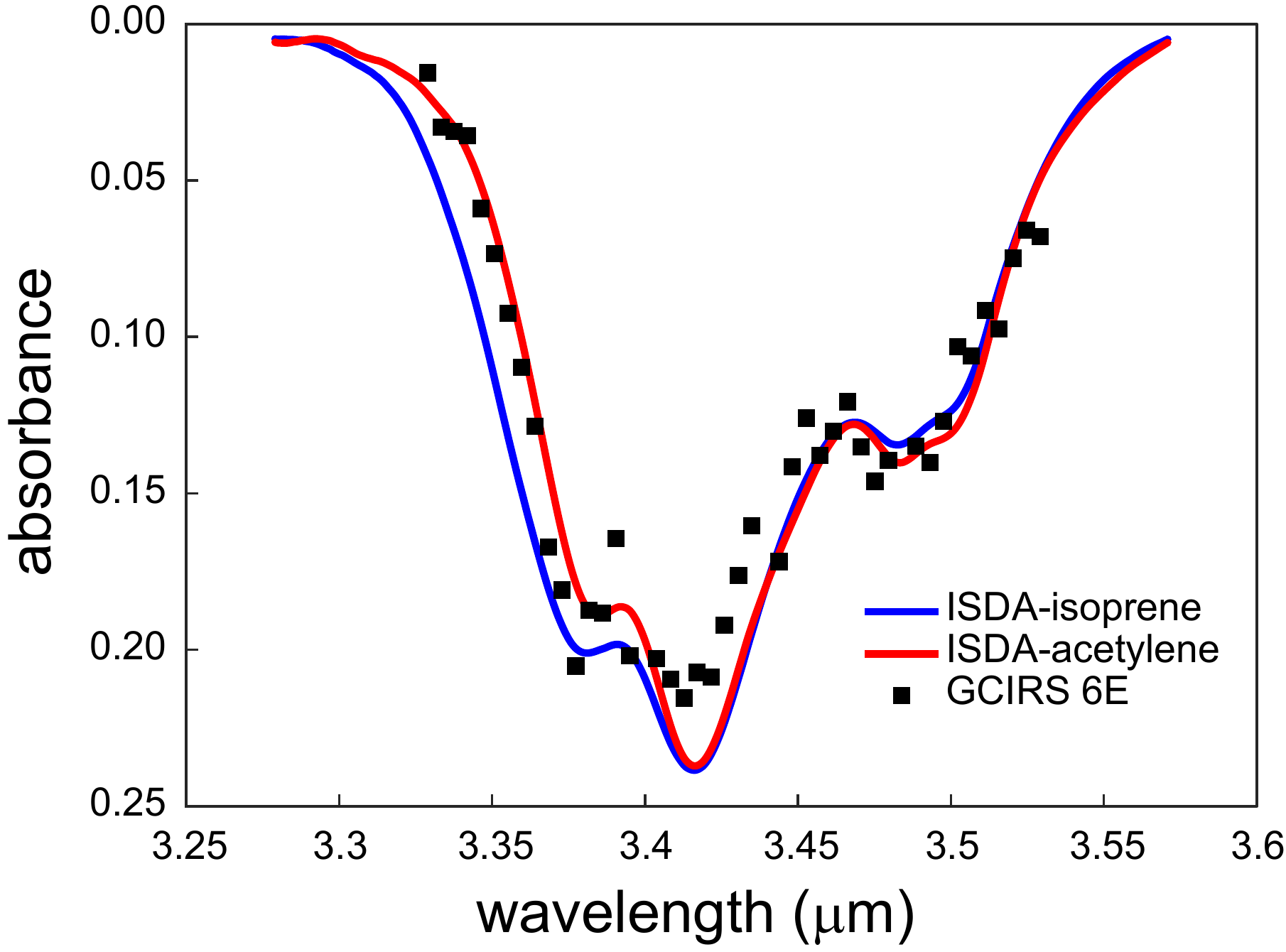}
       \caption{A comparison of the ISM absorption spectrum through the line of sight of the Galactic Centre source; GCIRS 6E \citep{Pendleton1994} and the aliphatic absorption feature of ISDA-isoprene and ISDA-acetylene. A normalising factor has been applied to facilitate comparison of the absorption profiles. }
     \label{fig4}
\end{figure}

Averaged spectra were obtained using repeated FTIR measurements with different amounts of sample, normalised in the 3.4\,$\mu$m region. The aliphatic absorption feature of ISDA-isoprene and ISDA-acetylene are compared with observational spectra (through the line of sight of the Galactic Centre source; GCIRS 6E) from \cite{Pendleton1994} in Figure \ref{fig4}, indicating that the results are in remarkably good agreement, and are promising ISDAs. The complex profile and sub-peak positions of the 3.4\,$\mu$m aliphatic C-H stretch absorption were found to be good in agreement with the profile of the 3.4\,$\mu$m absorption feature in the interstellar spectra.

The symmetric C-H stretches of CH$_3$ and CH$_2$ are respectively found at 3.48 and 3.50\,$\mu$m, and the asymmetric stretches are found at 3.38 and 3.42\,$\mu$m. Any tertiary C-H stretches are found near 3.44\,$\mu$m. No attempts were made to deconvolve the observed spectrum into sub-peaks, in keeping with our holistic approach.

\subsection{Solid-state $^{13}$C NMR Measurements}
The NMR technique requires a large sample, and therefore long times are needed to produce sufficient sample under simulated circumstellar/interstellar medium conditions. Our apparatus generated 14.80\,mg ISDA-isoprene and 7.67\,mg ISDA-acetylene in the available time.

\begin{figure}
  \leavevmode
    \centering
      \epsfxsize= 8cm
      \epsfbox{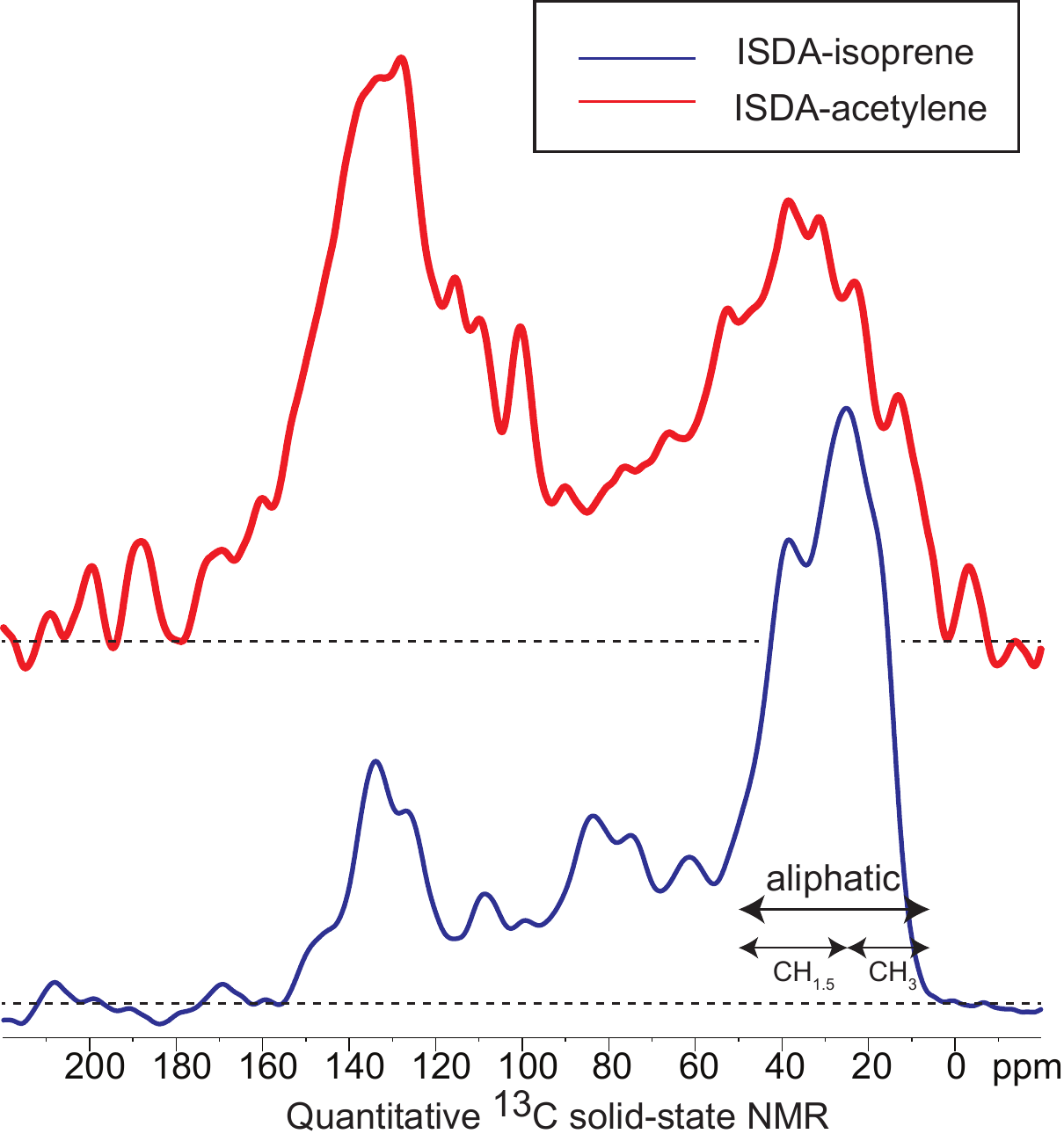}
       \caption{The quantitative solid-state $^{13}$C NMR spectra of the ISDA-acetylene and ISDA-isoprene. The ordinate is proportional to $^{13}$C content.}
     \label{fig7}
\end{figure}

The quantitative solid-state $^{13}$C NMR spectra of ISDA$-$acetylene and ISDA$-$isoprene are plotted in Figure \ref{fig7}. The samples show $^{13}$C NMR signal ranging between 6\,ppm to 150\,ppm (parts per million shift from tetramethylsilane). The signal of the aliphatic CH, CH$_{2}$ and CH$_{3}$ carbons are located in the region $6 - 50$\,ppm, with the CH$_3$ signal at the lower end ($6-25$\,ppm). The region $25-50$\,ppm is assigned by $^{13}$C NMR spectroscopists to `CH$_{1.5}$', since the CH and CH$_2$ contributions cannot be easily distinguished. In the region $50-90$\,ppm, sp$^3$ carbon bound to oxygen is detected. The region $90-150$\,ppm is assigned to sp$^{2}$ carbon.

The weight ratio, $r_C$, of aliphatic carbon in the interstellar dust analogues was determined using an external spin-counting reference (adamantane). We were not able to separate aliphatic carbons into CH$_{2}$ and CH$_{3}$ groups owing to insufficient spectral resolution.

The composition of ISDAs obtained by quantitative solid-state $^{13}$C NMR results are summarised in Table \ref{tab:1}. The aliphatic carbon (CH, CH$_{2}$ and CH$_{3}$) weight ratio was found to be 29.5\,\% and 14.5\,\% for ISDA-isoprene and ISDA-acetylene, respectively, corresponding to 57\,\% and 38\,\% of the total carbon. ISDA-acetylene contains far more sp$^2$ carbon, with sp$^{2}$/sp$^{3}=0.36$ and 0.81 for ISDA-isoprene and ISDA-acetylene, respectively. The shortfall between aliphatic and sp$^3$ carbon is due to oxygen-bearing carbon (alcohols and ethers). These are in a low protonation state, and as such are not expected to significantly contribute to the FTIR absorption. No sp-hybridised carbon was observed in the samples, despite acetylene's native character.

\begin{table}
  \caption{ISDA Composition from Quantitative solid-state $^{13}$C NMR}
  \label{tab:1}
   \centering
  \begin{tabular}{@{}lll@{}}
  \hline
  \hline
&  ISDA-isoprene & ISDA-acetylene  \\
& (weight $\%$) & (weight $\%$) \\
 \hline
Aliphatic Carbon  		& 29.5 	&14.5\\
(CH$_{2}$ and CH$_{3}$) & 	& \\
 \hline
sp$^{2}$ Carbon  		&13.4 	&16.9\\
sp$^{3}$ Carbon 		&38.0 	&21.0 \\
 \hline
Total Carbon 				&51.5	&37.9\\
\hline
\end{tabular}
\end{table}

\subsection{Measurements of Integrated Absorption Coefficient ($A$) at 3.4\,$\mu$m}
In our holistic approach, we aim to obtain the total integrated absorption coefficient for the 3.4\,$\mu$m aliphatic feature of the ISDAs, without distinguishing the CH, CH$_2$ or CH$_3$ groups. That the infrared spectra in Figure \ref{fig4} are so similar to the extinction towards the galactic centre shows that the CH$_2$/CH$_3$ ratio in the ISDAs is close to astronomical.

\begin{figure}
  \leavevmode
    \centering
      \epsfxsize= 8cm
      \epsfbox{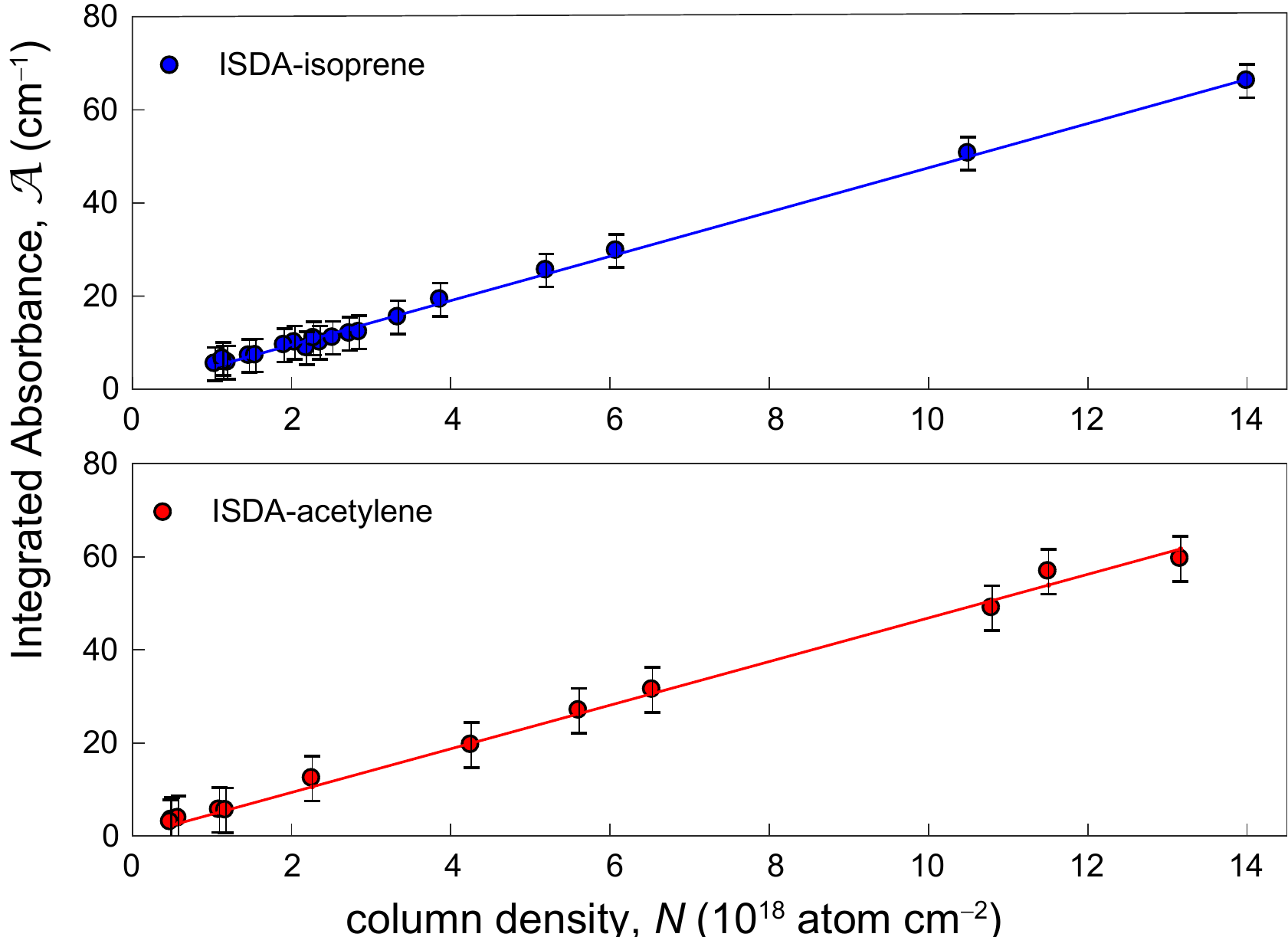}
       \caption{The integrated absorbance ($\mathcal{A}$) (cm$^{-1}$) as a function of column density (atom cm$^{-2}$) for ISDA-isoprene (upper panel) and ISDA-acetylene (lower panel). The integrated absorption coefficients ($A$) (cm atom$^{-1}$) were obtained from the slope of the linear fit. }
     \label{fig8}
\end{figure}

The integrated absorbance ($\mathcal{A}$, cm$^{-1}$) is plotted as a function of aliphatic carbon column density ($N$, group\,cm$^{-2}$) in Figure \ref{fig8}. The integrated absorption coefficients ($A$, cm group$^{-1}$) were obtained from the slope of the linear fit. The absorption coefficients obtained for ISDAs are $4.76(8)\times10^{-18}$\,cm\,group$^{-1}$ for ISDA-isoprene and $4.69(14)\times10^{-18}$\,cm\,group$^{-1}$ for ISDA-acetylene. These results, along with peak cross sections ($\sigma$, cm$^2$\,group$^{-1}$) and mass extinction coefficient ($\kappa$, cm$^2$\,g$^{-1}$) are given in Table \ref{tab:x}. These quantities are related by

\[
A = \int \sigma(\bar{\nu}) d\bar{\nu}
\]
and
\[
\kappa = \sigma/m_{CH_x},
\]
where $m_{CH_x}$ is the average mass of an aliphatic group. Within error, the absorption coefficients of the ISDAs are identical.

The absorption coefficients are compared with the literature \citep{Sandford1991, Dartois2007} in Table \ref{tab:x}. Since we were not able to separate aliphatic carbons into CH$_{2}$ and CH$_{3}$ groups, we could not calculate \textit{A}$_{CH2}$ and \textit{A}$_{CH3}$ separately. Therefore we indicated \textit{A}$_{CHx}$ and assumed $N(CH_2)/N(CH_3)=2$ for comparison in Table \ref{tab:x} (the comparison is not very sensitive to this ratio).

The results obtained in this study are less than half those obtained by \citet{Sandford1991} and \citet{Dartois2004} using small hydrocarbons. They are, however, consistent with the work of \citet{Duley1998}, who analysed a dust analogue. In converting our values into mass extinction coefficients, accounting for only the aliphatic carbon mass, we obtain values consistent with the previous studies of \citet{Furton1999}, \citet{Mennella2002} and \citet{Gadallah2015}. It has been noted by \citet{Steglich2013} that aliphatic material in dust seems to have a lower extinction coefficient than in small molecules. Our studies have rigorously confirmed this.

Armed with a reliable extinction coefficient, below we assess the astrophysical implications.

\begin{table*}
\footnotesize
 \begin{center}
  \caption{Column densities ($10^{18}$cm$^{-2}$) and aliphatic carbon abundances (ppm) towards Galactic Centre sources.
 Normalised carbon abundances were calculated using $N(H) = 2.04\times10^{21}$\,cm$^{-2}$\,mag$^{-1}$\,H \citep{Zhu2017} (A$_{V}\sim$30).}
\centering
  \label{tab:7}
  \begin{tabular}{ccccccccccc}
    \hline
 & \textbf{GC Sources} & \textbf{IRS 1W} & \textbf{IRS 3} & \textbf{IRS 6E} &  \textbf{IRS 7} &  \textbf{IRS 8} &   \textbf{IRS 12N} &  \textbf{IRS 19} & \textbf{Average} &\textbf{IRS 6E}$^{a}$\\
   \hline
  &\textbf{ $\tau_{3.4\,\mu m}$ } & 0.179 &0.310	&0.259 	& 0.147	& 0.258	 & 0.357 & 0.247 & 0.220 & ... \\
    \hline
&\textbf{$\mathcal{A}$/cm$^{-1}$}& 19.424	& 33.640	& 28.105	& 15.952	& 27.997	& 38.740	& 23.873 &26.819	& 23.372\\
  \hline
 \textbf{ISDA-iso.} & $N/10^{18}$\,cm$^{-2}$ & 4.13	& 7.16& 5.98 & 3.39 & 5.96 & 8.24 & 5.08  & 5.71& 4.97\\
 & ppm aliphatic &68	&117	& 98	& 55	& 97	& 135	& 83 & 93& 81\\
  \hline
\textbf{ISDA-ac.}  & $N/10^{18}$\,cm$^{-2}$ & 4.05 & 7.01 & 5.86 &	3.32 & 5.83 	& 8.07 & 4.97 & 5.59 & 4.87\\
 & ppm aliphatic &66	 & 115	& 96	& 54	& 95	& 132	& 81 & 91	& 80\\
\hline
\end{tabular}
    \end{center}
    \begin{flushleft}
\begin{footnotesize}
 $^{a}$$\mathcal{A}$ obtained by integration of the 3.4\,$\mu$m optical thickness spectra presented in Figure 4 \citep{Pendleton1994}.
\end{footnotesize}
\end{flushleft}
\end{table*}

\subsection{Astrophysical Implications}\label{Section4}
\subsubsection{Aliphatic Carbon Column Densities}
Using the absorption coefficients determined above, the column density of aliphatic carbon was calculated for lines of sight toward Galactic Centre sources. Since the results for both ISDAs are so similar, only ISDA-acetylene results are discussed here, but the results from both ISDAs are reported in Table \ref{tab:7}.

\citet{Pendleton1994} studied the 3.4\,\,$\mu$m feature towards a range of sources including the galactic centre and local diffuse ISM. Towards GC IRS 6E at $A_V=31$, they estimated a total carbon column density of $2.2\times10^{19}$\,cm$^{-2}$ from $N(H)=1.9\times10^{21}A_V$ and $N(C)/N(H)=370$\,ppm. A total aliphatic column density of $9.3\times10^{17}$ was reported, comprising 4.2\,\% of the available carbon. In the light of the present study it appears that this is significantly underestimated.

\citet{Pendleton1994} recorded $\tau$ for various sub-features attributed to CH$_2$ and CH$_3$ groups, and then used the peak widths and integrated absorbances of small molecules \citep{Sandford1991,Hendecourt1986} to calculate the column densities. In this study, we treat the aliphatic absorption as a single feature and have determined its integrated absorption coefficient to be $4.7\times10^{-18}$\,cm\,group$^{-1}$. Moreover, the equivalent width of our feature is 111\,cm$^{-1}$ (integral of the feature with unit peak absorbance), which compares favorably with the integral of the feature towards GC IRS 6E ($\sim108$\,cm$^{-1}$). The widths of the small molecule features from \cite{Sandford1991} are much smaller, $\sim20$\,cm$^{-1}$. The larger $A$ values and the smaller $\Delta\bar{\nu}$ values compound to underestimate the column density of aliphatic carbon. Reappraising the observations of \cite{Pendleton1994}, we determine an aliphatic carbon column density of $4.87\times10^{18}$\,cm$^{-2}$. This is a factor of five higher than previously reported, corresponding to about 22\,\% of the available carbon. We recommend that the total aliphatic carbon column densities of \cite{Pendleton1994} be increased by a factor of 5.2, which increases the percentage of aliphatic carbon in the local diffuse ISM to about 10\,\%.

With the $\tau_{3.4\,\mu m}$ values obtained from \cite{Chiar2002}, and the equivalent width of $\Delta\bar{\nu}=108.5$\,cm$^{-1}$ obtained from integration of the 3.4\,$\mu$m feature from \citep{Pendleton1994}, we have calculated the aliphatic carbon column densities towards a range of sources, which are summarised in Table \ref{tab:7}. Employing the $\tau_{3.4\,\mu m}$ value of GC IRS 6E from \citet{Chiar2002}, the aliphatic carbon column density was found to be $5.86\times10^{18}$\,cm$^{-2}$.

\subsubsection{Aliphatic Carbon Abundance}
From the H column densities calculated by gas-to-extinction ratio recently published by \cite{Zhu2017}, we obtained normalised aliphatic carbon abundances. Normalised carbon abundances (C/H) (ppm) were calculated based on gas-to-extinction ratio $N(H) = 2.04 \times 10^{21}$\,cm$^{-2}A_V$ \citep{Zhu2017}, assuming $A_{V}\sim30$. The resultant aliphatic and total carbon column densities and normalised abundances for the line of sights through the Galactic Centre are compared in Table \ref{tab:7}. The aliphatic abundance was found to be in the range $54-132$\,ppm. If we use the gas-to-extinction ratio from \cite{Bohlin1978} to obtain the H column density, the normalised carbon abundance values increase slightly.

There are inhomogeneities in the aliphatic carbon abundances, even on small spatial scales in the GC based on the reported  $\tau_{3.4\,\mu m}$ values in Table \ref{tab:7}. However, our reported relative abundances are also due to the extinction. We assume $A_{V}\sim30$ mag and based on the fluctuations in value of $A_{V}$ towards each line of sight towards the GC, the normalised carbon abundances would be different to the values presented in Table \ref{tab:7}.

There are major uncertainties in  $\tau_{3.4\,\mu m}$ values arising from determination of transmitted flux ($I$) due to the resolution of the observational absorption spectra and initial flux ($I_0$) due to the estimation of the continuum (blackbody radiation curve of the background light source object) of the spectra.

\citealt{Moultaka2004} showed that $\tau_{3.4\,\mu m}$ values differ for each line of sight based on the continuum fit. For instance, they reported that $\tau_{3.4\,\mu m} = 0.49$ towards GCIRS 16C (the maximum $\tau_{3.4\,\mu m}$ reported towards the GC in the literature.) However, they highlighted a lower limit of $\tau_{3.4\,\mu m}$ = 0.14 towards the same source. Furthermore, this aspect together with spectral resolution causes discrepancies in the $\tau_{3.4\,\mu m}$ values reported in the literature for the same sight lines of the GC. For example, \cite{Chiar2002} reported that $\tau_{3.4\,\mu m} = 0.147$ towards GCIRS 7, whereas \citealt{Moultaka2004} reported $\tau_{3.4\,\mu m} = 0.41$. This shows that $\tau_{3.4\,\mu m}$ values depend strongly on the measurement and analysis techniques, giving rise to large uncertainties in carbon abundances. For more reliable information on the distribution of carbon in ISM dust towards the GC, we need to use $\tau_{3.4\,\mu m}$ values obtained by a reliable method through all the sight-lines of interest.

There is no reason to expect the total carbon abundance to be homogenous across all lines of sight in the ISM. There would be regions in the ISM where the total carbon abundance would exceed the cosmic abundances and in some regions where there would be lack of carbon. For instance, \cite{Pendleton1994} showed that $A_V/\tau_{3.4\,\mu m}$ is lower towards the galactic centre than in the local diffuse ISM. Therefore, to rigorously quantify interstellar carbon, we need to determine the total carbon abundances along many lines of sight.
In order to obtain reliable results, we need to directly measure carbon abundance in the dust and gas phases, rather than relying on estimations. One step towards this is obtaining rigorous integrated absorption coefficients for identifiable carbon, as in the present study.

Importantly, our reported aliphatic abundances, though higher than those obtained using small molecule absorption coefficients, do not break the carbon budget. We report $54-135$\,ppm aliphatic carbon towards GC sources. The ISM dust carbon abundances obtained from the atmospheres of young F, G type stars \citep{Sofia2001} is about 220 \,ppm. Grain-size distribution models (e.g. \citealt{Mathis1977} or  \citealt{Kim1996}) require around 300 \,ppm carbon, and more recent models such as that of \cite{Zubko2004} and \cite{Li2001} propose around 250\,ppm carbon in dust \citep{Dwek2004}. As such, the values reported here suggest that a significant component of the dust is aliphatic, but that there is plenty of carbon available for aromatic and olefinic structures such as the 2175\,\AA\ carrier.

Future laboratory studies considering sp$^{2}$ carbon content of the ISDAs which can reproduce the strength and shape of the UV Bump would give the opportunity to measure the amount of sp$^{2}$ carbon and therefore the total carbon (sp+sp$^{2}$+sp$^{3}$) abundance would be measured more precisely.

\section{Conclusion}\label{Section5}
The interstellar dust analogues produced in the laboratory enable us to better understand the nature of the dust particles in the ISM. In this study, we produced reliable dust analogues from gas phase precursor molecules by mimicking interstellar/circumstellar conditions. Ensuring that the laboratory spectra of ISDAs matched interstellar spectra, we calculated the 3.4\,$\mu$m absorption coefficients by combining FTIR and $^{13}C$ NMR spectroscopy. Despite the chemical difference, the absorption coefficients obtained for both interstellar dust analogues were found to be in close agreement, supporting the reliability of our method and the interstellar dust analogues studied here.

The 3.4\,$\mu$m aliphatic carbon integrated absorption coefficients obtained in this study are lower than the values reported in the literature based on small molecules, but consistent with other studies on interstellar dust analogues. Given that the absorption profiles of our interstellar dust analogues matched that obtained from ISM observations, we are confident that our values are reliable.

We determined the aliphatic carbon abundance in ISM dust towards the Galactic Centre using our 3.4\,$\mu$m integrated absorption coefficients and the $\tau_{3.4\,\mu m}$ values from the literature. The resultant aliphatic carbon column densities are least five times higher than some values reported previously. Using the two ISDA integrated absorption coefficients, we obtained an abundance range between $\sim54 - 135$\,ppm for aliphatic carbon in the ISM. This leaves a substantial proportion of the dust-bound carbon to be found in aromatic or olefinic structures.

\section*{Appendix}

The optical depth of the 3.4 $\mu$m absorption feature is related to the number of aliphatic carbon groups along the line of sight

\begin{equation} \label{eq:1}
N =\frac{\tau \Delta \bar{\nu}} {\textit{A}}
\end{equation}

Therefore, the optical depth, $\tau$, of the 3.4\,$\mu$m absorption feature can be used to determine the number of aliphatic groups by using the integrated absorption coefficient, $A$ (cm\,group$^{-1}$), and the full width at half maximum of the feature, $\Delta \bar{\nu}$ (cm$^{-1}$), which is the same as the equivalent width for triangular absorption features.

Transmittance ,$T$, is the fraction of incident electromagnetic power that is transmitted through a material. The ratio of the intensity of the transmitted flux, $I$, to the initial flux, $I_0$, gives the transmittance.
	
Depending on whether a decadic or natural logarithm is used, transmittance is related to optical depth ($\tau = A_e$) or decadic absorbance $A_{10}$,
\begin{equation} \label{eq:2}
T =\frac{I}{I_0} = \exp(-\tau) = \exp(-\rm A_{e}) = 10^{-\rm A_{10}}
\end{equation}

In astronomy, the optical depth, $\tau$, is used instead of absorbance, $A_e=-\log_e T$. In chemistry, absorbance is the decadic logarithm of the ratio of incident to the transmitted radiant power through a material (A$_{10}$). Therefore,
	
\begin{equation} \label{eq:3}
\tau = \log_e 10\times A_{10}
\end{equation}

The Beer-Lambert Law is often stated $A_{10}=\epsilon c l$, where $\epsilon$ is the decadic molar extinction coefficient, $c$ is the concentration of the absorbing species and $l$ is the pathlength. The astronomical version of this law is $\tau = \sigma N$, where $\sigma$ is the absorption cross section and $N$ is the column density. The cross section $\sigma$ is a function of wavenumber, $\bar{\nu}$ (cm$^{-1}$) and the integrated absorption coefficient (integrated cross-section) $A$ is given by the integral over the entire absorption feature, with units of cm group$^{-1}$.

\begin{equation} \label{eq:13}
A = \int \sigma({\bar{\nu}}) d\bar{\nu} = \frac{1}{N}\int \tau d\bar{\nu} = \frac{\mathcal{A}}{N}
\end{equation}	

A plot of $\mathcal{A}=\int \tau d\bar{\nu}$ as a function of $N$ is thus a straight line with slope $A$. The column densities in our samples were obtained from the aliphatic carbon ratio, $r_C$ of the analogue measured by $^{13}$C NMR spectroscopy.

\begin{equation} \label{eq:16}
{\textit{N}} = N_A \frac{r_C m_a}{M_C s_p}
\end{equation}	

where mass of the analogue in the pellet is $m_a$, the molar mass of carbon is $M_C=12.01$\,g\,mol$^{-1}$, the surface area of the pellet is $s_p$ and $N_A=6.022\times10^{23}$\,mol$^{-1}$ is Avogadro's number.

\section*{Acknowledgments}
We would like to thank Drs Yvonne Pendleton and Emmanuel Dartois for their support and for supplying the observational data. B.G. would like to thank to The Scientific and Technological Research Council of Turkey (T\"{U}B\.{I}TAK) as our work has been supported with 2214/A International Research Fellowship Programme. We would like to thank Dr Simon Hager for technical assistance and use of facilities at the Electron Microscope Unit at UNSW. We would like to thank Alireza Kharazmi for his support for FTIR studies.

T.W.S. acknowledges the Australian Research Council for a Future Fellowship (FT130100177).  This work was supported by the Australian Research Council Centre of Excellence in Exciton Science (CE170100026).

 \bibliographystyle{mn2e}{}
 \bibliography{List}

\begin{thebibliography}{66}
\providecommand{\natexlab}[1]{#1}

\bibitem[{{Adamson} et~al.(1990){Adamson}, {Whittet} \& {Duley}}]{Adamson1990}
{Adamson} A.~J., {Whittet} D.~C.~B., {Duley} W.~W., 1990, MNRAS, 243, 400

\bibitem[{{Asplund} et~al.(2006){Asplund}, {Grevesse} \& {Jacques
  Sauval}}]{Asplund2005}
{Asplund} M., {Grevesse} N., {Jacques Sauval} A., 2006, Nucl. Phys. A, 777, 1

\bibitem[{{Asplund} et~al.(2009){Asplund}, {Grevesse}, {Sauval} \&
  {Scott}}]{Asplund2009}
{Asplund} M., {Grevesse} N., {Sauval} A.~J., {Scott} P., 2009, ARA\&A, 47, 481

\bibitem[{{Bohlin} et~al.(1978){Bohlin}, {Savage} \& {Drake}}]{Bohlin1978}
{Bohlin} R.~C., {Savage} B.~D., {Drake} J.~F., 1978, ApJ, 224, 132

\bibitem[{{Butchart} et~al.(1986){Butchart}, {McFadzean}, {Whittet}, {Geballe}
  \& {Greenberg}}]{Butchart1986}
{Butchart} I., {McFadzean} A.~D., {Whittet} D.~C.~B., {Geballe} T.~R.,
  {Greenberg} J.~M., 1986, A\&A, 154, L5

\bibitem[{{Cardelli} et~al.(1989){Cardelli}, {Clayton} \&
  {Mathis}}]{Cardelli1989}
{Cardelli} J.~A., {Clayton} G.~C., {Mathis} J.~S., 1989, ApJ, 345, 245

\bibitem[{{Cardelli} et~al.(1996){Cardelli}, {Meyer}, {Jura} \&
  {Savage}}]{Cardelli1996}
{Cardelli} J.~A., {Meyer} D.~M., {Jura} M., {Savage} B.~D., 1996, ApJ, 467, 334

\bibitem[{{Cartledge} et~al.(2004){Cartledge}, {Lauroesch}, {Meyer} \&
  {Sofia}}]{Cartledge2004}
{Cartledge} S.~I.~B., {Lauroesch} J.~T., {Meyer} D.~M., {Sofia} U.~J., 2004,
  ApJ, 613, 1037

\bibitem[{{Cartledge} et~al.(2006){Cartledge}, {Lauroesch}, {Meyer} \&
  {Sofia}}]{Cartledge2006}
{Cartledge} S.~I.~B., {Lauroesch} J.~T., {Meyer} D.~M., {Sofia} U.~J., 2006,
  ApJ, 641, 327

\bibitem[{Chiar et~al.(1998)Chiar, Pendleton, Geballe \& Tielens}]{Chiar1998}
Chiar J.~E., Pendleton Y.~J., Geballe T.~R., Tielens A.~G.~G.~M., 1998, ApJ,
  507, 281

\bibitem[{{Chiar} et~al.(2002){Chiar}, {Adamson}, {Pendleton}, {Whittet},
  {Caldwell} \& {Gibb}}]{Chiar2002}
{Chiar} J.~E., {Adamson} A.~J., {Pendleton} Y.~J., {Whittet} D.~C.~B.,
  {Caldwell} D.~A., {Gibb} E.~L., 2002, ApJ, 570, 198

\bibitem[{{Chiar} et~al.(2013){Chiar}, {Tielens}, {Adamson} \&
  {Ricca}}]{Chiar2013}
{Chiar} J.~E., {Tielens} A.~G.~G.~M., {Adamson} A.~J., {Ricca} A., 2013, ApJ,
  770, 78

\bibitem[{{Chiar} et~al.(2000)}]{Chiar2000}
{Chiar} J.~E., {Tielens} A.~G.~G.~M., {Whittet} D.~C.~B., {Schutte} W.~A.,
  {Boogert} A.~C.~A., {Lutz} D., {van Dishoeck} E.~F., {Bernstein} M.~P., 2000,
  ApJ, 537, 749

\bibitem[{Contreras \& Salama(2013)}]{Contreras2013}
Contreras C.~S., Salama F., 2013, ApJS, 208, 6

\bibitem[{{Dartois} et~al.(2004){Dartois}, {Marco}, {Mu{\~n}oz-Caro}, {Brooks},
  {Deboffle} \& {d'}Hendecourt}]{Dartois2004}
{Dartois} E., {Marco} O., {Mu{\~n}oz-Caro} G.~M., {Brooks} K., {Deboffle} D.,
  {d'}Hendecourt L., 2004, A\&A, 423, 549

\bibitem[{{Dartois, E.} et~al.(2007)}]{Dartois2007}
{Dartois, E.} et~al., 2007, A\&A, 463, 635

\bibitem[{{d'Hendecourt} \& {Allamandola}(1986)}]{Hendecourt1986}
{d'Hendecourt} L.~B., {Allamandola} L.~J., 1986, A\&ASS, 64, 453

\bibitem[{{Diplas} \& {Savage}(1994)}]{Diplas1994}
{Diplas} A., {Savage} B.~D., 1994, ApJS, 93, 211

\bibitem[{Duley \& Williams(1981)}]{Duley1981}
Duley W.~W., Williams D.~A., 1981, MNRAS, 196, 269

\bibitem[{{Duley} et~al.(1998){Duley}, {Scott}, {Seahra} \&
  {Dadswell}}]{Duley1998}
{Duley} W.~W., {Scott} A.~D., {Seahra} S., {Dadswell} G., 1998, ApJL, 503, L183

\bibitem[{Dwek(1997)}]{Dwek1997}
Dwek E., 1997, ApJ, 484, 779

\bibitem[{{Dwek}(2005)}]{Dwek2004}
{Dwek} E., 2005, in C.C. {Popescu}, R.J. {Tuffs}, eds, The Spectral Energy
  Distributions of Gas-Rich Galaxies: Confronting Models with Data. American
  Institute of Physics Conference Series, Vol. 761, pp. 103--122

\bibitem[{{Ehrenfreund} et~al.(1991){Ehrenfreund}, {Robert}, {D'Hendecourt} \&
  {Behar}}]{Ehrenfreund1991}
{Ehrenfreund} P., {Robert} F., {D'Hendecourt} L., {Behar} F., 1991, A\&A, 252,
  712

\bibitem[{{Fitzpatrick}(1999)}]{Fitzpatrick1999}
{Fitzpatrick} E.~L., 1999, PASP, 111, 63

\bibitem[{Fitzpatrick \& Massa(2007)}]{Fitzpatrick2007}
Fitzpatrick E.~L., Massa D., 2007, ApJ, 663, 320

\bibitem[{Furton et~al.(1999)Furton, Laiho \& Witt}]{Furton1999}
Furton D.~G., Laiho J.~W., Witt A.~N., 1999, ApJ, 526, 752

\bibitem[{Gadallah(2015)}]{Gadallah2015}
Gadallah K.~A.~K., 2015, Adv. Space Res., 55, 705

\bibitem[{{Geballe} et~al.(2009){Geballe}, {Mason}, {Rodr{\'{\i}}guez-Ardila}
  \& {Axon}}]{Geballe2009}
{Geballe} T.~R., {Mason} R.~E., {Rodr{\'{\i}}guez-Ardila} A., {Axon} D.~J.,
  2009, ApJ, 701, 1710

\bibitem[{Grevesse \& Sauval(1998)}]{Grevesse1998}
Grevesse N., Sauval A., 1998, Space Sci. Rev., 85, 161

\bibitem[{Henning et~al.(2004)Henning, J{\"{a}}ger \& Mutschke}]{Henning2004}
Henning T., J{\"{a}}ger C., Mutschke H., 2004, Astrophysics of Dust, 309, 603

\bibitem[{{Imanishi}(2000)}]{Imanishi2000}
{Imanishi} M., 2000, MNRAS, 319, 331

\bibitem[{{Kim} \& {Martin}(1996)}]{Kim1996}
{Kim} S.~H., {Martin} P.~G., 1996, ApJ, 462, 296

\bibitem[{Kova{\v{c}}evi{\'{c}} et~al.(2005)Kova{\v{c}}evi{\'{c}},
  Stefanovi{\'{c}}, Berndt, Pendleton \& Winter}]{Kovacevic2005}
Kova{\v{c}}evi{\'{c}} E., Stefanovi{\'{c}} I., Berndt J., Pendleton Y.~J.,
  Winter J., 2005, ApJ, 623, 242

\bibitem[{Kwok(2009)}]{Kwok2009}
Kwok S., 2009, Astrophys. Space Sci., 319, 5

\bibitem[{{Lee} \& {Wdowiak}(1993)}]{Lee1993}
{Lee} W., {Wdowiak} T.~J., 1993, ApJL, 417, L49

\bibitem[{{Lequeux} \& {Jourdain de Muizon}(1990)}]{Lequeux1990}
{Lequeux} J., {Jourdain de Muizon} M., 1990, A\&A, 240, L19

\bibitem[{Li \& Draine(2001)}]{Li2001}
Li A., Draine B.~T., 2001, ApJ, 554, 778

\bibitem[{Lodders(2003)}]{Lodders2003}
Lodders K., 2003, ApJ, 591, 1220

\bibitem[{Mason et~al.(2004)Mason, Wright, Pendleton \& Adamson}]{Mason2004}
Mason R.~E., Wright G., Pendleton Y., Adamson A., 2004, ApJ, 613, 770

\bibitem[{{Mathis}(1994)}]{Mathis1994}
{Mathis} J.~S., 1994, ApJ, 422, 176

\bibitem[{Mathis et~al.(1977)Mathis, Rumpl \& Nordsieck}]{Mathis1977}
Mathis J.~S., Rumpl W., Nordsieck K.~H., 1977, ApJ, 217, 425

\bibitem[{Matrajt et~al.(2005)Matrajt, {Mu{\~{n}}oz Caro}, Dartois,
  D'Hendecourt, Deboffle \& Borg}]{Matrajt2005}
Matrajt G., {Mu{\~{n}}oz Caro} G.~M., Dartois E., D'Hendecourt L., Deboffle D.,
  Borg J., 2005, A\&A, 433, 979

\bibitem[{McFadzean et~al.(1989)McFadzean, Whittet, Bode, Adamson \&
  Longmore}]{McFadzean1989}
McFadzean A.~D., Whittet D.~C.~B., Bode M.~F., Adamson A.~J., Longmore A.~J.,
  1989, MNRAS, 241, 873

\bibitem[{{Mennella} et~al.(1999){Mennella}, {Brucato}, {Colangeli} \&
  {Palumbo}}]{Mennella1999}
{Mennella} V., {Brucato} J.~R., {Colangeli} L., {Palumbo} P., 1999, ApJL, 524,
  L71

\bibitem[{Mennella et~al.(2002)Mennella, Brucato, Colangeli \&
  Palumbo}]{Mennella2002}
Mennella V., Brucato J.~R., Colangeli L., Palumbo P., 2002, ApJ, 569, 531

\bibitem[{{Mishra} \& {Li}(2017)}]{Mishra2017}
{Mishra} A., {Li} A., 2017, ApJ, 850, 138

\bibitem[{Moultaka et~al.(2004)Moultaka, Eckart, Viehmann, Mouawad,
  Straubmeier, Ott \& Sch{\"{o}}del}]{Moultaka2004}
Moultaka J., Eckart A., Viehmann T., Mouawad N., Straubmeier C., Ott T.,
  Sch{\"{o}}del R., 2004, A\&A, 425, 529

\bibitem[{{Mu{\~n}oz Caro} et~al.(2008){Mu{\~n}oz Caro}, {Dartois} \&
  {Nakamura-Messenger}}]{MunozCaro2008}
{Mu{\~n}oz Caro} G.~M., {Dartois} E., {Nakamura-Messenger} K., 2008, A\&A, 485,
  743

\bibitem[{Parvathi et~al.(2012)Parvathi, Sofia, Murthy \& Babu}]{Parvathi2012}
Parvathi V.~S., Sofia U.~J., Murthy J., Babu B.~R.~S., 2012, ApJ, 760, 36

\bibitem[{{Pendleton} et~al.(1994){Pendleton}, {Sandford}, {Allamandola},
  {Tielens} \& {Sellgren}}]{Pendleton1994}
{Pendleton} Y.~J., {Sandford} S.~A., {Allamandola} L.~J., {Tielens}
  A.~G.~G.~M., {Sellgren} K., 1994, ApJ, 437, 683

\bibitem[{{Przybilla} et~al.(2008){Przybilla}, {Nieva} \&
  {Butler}}]{Przybilla2008}
{Przybilla} N., {Nieva} M.~F., {Butler} K., 2008, ApJL, 688, L103

\bibitem[{Robertson(2002)}]{Robertson2002}
Robertson J., 2002, Mater. Sci. Eng. R, 37, 129

\bibitem[{{Sandford} et~al.(1991){Sandford}, {Allamandola}, {Tielens},
  {Sellgren}, {Tapia} \& {Pendleton}}]{Sandford1991}
{Sandford} S.~A., {Allamandola} L.~J., {Tielens} A.~G.~G.~M., {Sellgren} K.,
  {Tapia} M., {Pendleton} Y., 1991, ApJ, 371, 607

\bibitem[{{Sandford} et~al.(2006)}]{Sandford2006}
{Sandford} S.~A. et~al., 2006, Science, 314, 1720

\bibitem[{{Schnaiter} et~al.(1999){Schnaiter}, {Henning}, {Mutschke}, {Kohn},
  {Ehbrecht} \& {Huisken}}]{Schnaiter1999}
{Schnaiter} M., {Henning} T., {Mutschke} H., {Kohn} B., {Ehbrecht} M.,
  {Huisken} F., 1999, ApJ, 519, 687

\bibitem[{Snow \& Witt(1995)}]{Snow1995}
Snow T.~P., Witt A.~N., 1995, Science, 270, 1455

\bibitem[{{Sofia} \& {Meyer}(2001)}]{Sofia2001}
{Sofia} U.~J., {Meyer} D.~M., 2001, ApJ, 554, L221

\bibitem[{{Stecher}(1965)}]{Stecher1965}
{Stecher} T.~P., 1965, ApJ, 142, 1683

\bibitem[{{Steglich} et~al.(2013)}]{Steglich2013}
{Steglich} M., {J{\"a}ger} C., {Huisken} F., {Friedrich} M., {Plass} W.,
  {R{\"a}der} H.~J., {M{\"u}llen} K., {Henning} T., 2013, ApJS, 208, 26

\bibitem[{Tielens et~al.(1996)Tielens, Wooden, Allamandola, Bregman \&
  Witteborn}]{Tielens1996}
Tielens A., Wooden D.~H., Allamandola L.~J., Bregman J., Witteborn F.~C., 1996,
  ApJ, 461, 210

\bibitem[{van Dishoeck(2014)}]{VanDishoeck2014}
van Dishoeck E.~F., 2014, Faraday Disc., 168, 9

\bibitem[{Whittet et~al.(1997)}]{Whittet1997}
Whittet D.~C.~B. et~al., 1997, ApJ, 490, 729

\bibitem[{{Wickramasinghe} \& {Allen}(1980)}]{Wickramasinghe1980}
{Wickramasinghe} D.~T., {Allen} D.~A., 1980, Nature, 287, 518

\bibitem[{Willner et~al.(1979)Willner, Russell, Puetter, Soifer \&
  Harvey}]{Willner1979}
Willner S.~P., Russell R.~W., Puetter R.~C., Soifer B.~T., Harvey P.~M., 1979,
  AJ, 229, L65

\bibitem[{{Zhu} et~al.(2017){Zhu}, {Tian}, {Li} \& {Zhang}}]{Zhu2017}
{Zhu} H., {Tian} W., {Li} A., {Zhang} M., 2017, MNRAS, 471, 3494

\bibitem[{Zubko et~al.(2004)Zubko, Dwek \& Arendt}]{Zubko2004}
Zubko V., Dwek E., Arendt R.~G., 2004, ApJS, 152, 211

\end{thebibliography}

\end{document}